\documentclass[%
 reprint,
 amsmath,amssymb,
 aps,
]{revtex4-2}

\usepackage{graphicx}
\usepackage{dcolumn}
\usepackage{bm}

\begin{document}

\preprint{APS/123-QED}

\title{The understanding of the penetration and clusterization of 1-alkanol in bilayer membrane: An open outlook based on atomistic molecular dynamics simulation}

\author{Anirban Polley}
\email{anirban.polley@gmail.com}
\affiliation{Shanmugha Arts, Science, Technology and Research Academy,
Tirumalaisamudram, Thanjavur, Tamilnadu 613401, India}




\date{\today}

\begin{abstract}
1-alkanols are well known to have anesthetic and penetration properties, though the mode of operation remains enigmatic. We perform extensive atomistic molecular dynamics simulation to study the penetration of 1-alkanols of different chain lengths in the dioleoyl-phosphatidylcholine (DOPC) bilayer model membrane. Our simulations show that the depth of penetration of 1-alkanol increases with chain length, $n$, and the deuterium order of the DOPC tail increases with the chain length of the acyl-chain of the 1-alkanol. We find a cut-off value for the length of the acyl-chain of 1-alkanol, $n = 12$, where 1-alkanol with a chain length greater than the cut-off value takes longer to penetrate the membrane. Our simulation study also demonstrates that the membrane exhibits clusters of 1-alkanols with acyl chains longer than the cut-off value, whereas 1-alkanols with acyl-chain shorter than the cut-off value are distributed homogeneously in the membrane and penetrate the membrane in a shorter time than longer-acyl-chain 1-alkanols.
These findings add to our understanding of the anomalies in anesthetic molecule partitioning in the cell membrane and may have implications for general anesthesia.

\end{abstract}

\maketitle


\section{\label{sec:level1}Introduction}

\noindent

General anesthesia \cite{seeman_anesthetic} is a well-known phenomenon, and anesthetics are widely used in hospitals for all painless surgical procedures. However, the molecular level of understanding in the mechanism of general anesthesia is not yet understood. It is still debatable whether general anesthesia is caused directly by the binding of anesthetics to specific proteins \cite{Frank_1994,Frank_1997} and blocking the protein's function by changing conformation, or whether it is caused indirectly by a lipid-mediated mechanism \cite{Regen_2009,Ueda_1998,Ueda_2001} while the anesthetics alter the membrane properties such as the total volume of the membrane, the volume occupied by the anesthetics within the membrane, the phase transition temperature of lipids, the lipid chain order, membrane thickness, or the lateral pressure profile of the membrane.

Apart from anesthesia, 1-alkanol has been well known as a penetration enhancer for the transdermal drug delivery. However, the mechanism of penetration enhancers is still unknown. It is interesting to know that 1-alkanol as anesthetics exhibits the `cutoff effect'. The anaesthetic potency of 1-alkanols is increased up to a certain cutoff chain length (dodecanol), and 1-alkanols with a chain length greater than the cutoff length have no anesthetic potency \cite{miller_anesthetic}.  

Several experimental studies have been conducted to explore the effect of the anesthetics on the properties of the lipid membrane, such as NMR spectroscopic studies that show ethanol molecules having a disordering effect on the lipids \cite{Feller_2002,Holte_1997,Barry_1995,Patra_2006,Joaquim_2011,Igor_2012} and X-ray studies that reveal ethanol having the potential to change the lipid membrane properties such as thickness above its main phase transition temperature \cite{Cantor_1997a,Cantor_1997b,Cantor_1998}. A few molecular dynamics studies have also been performed to study the influence of the anesthetics on the lipid membrane \cite{anirban_cpl13,Patra_2006,Bandyopadhyay_2004,Bandyopadhyay_2006,Smit_2004,
Terama_2008,Griepernau_2007,Faller_2007,
Vierl_1994,Jackson_2007,Gawrisch_1995,Dunn_1998,Mcintosh_1984}.

In the present work, using an atomistic molecular dynamics (MD) simulation technique, we investigate the penetration of 1-alkanols with varying chain lengths in a symmetric DOPC lipid bilayer membrane. 

The article is arranged as follows: we first discuss the details of the atomistic molecular dynamics simulations of the bilayer membrane. Next, we present our main results of the partitioning of 1-alkanols; the deuterium order parameter of DOPC acyl chain; the penetration time of 1-alkanols; the clusterization of 1-alkanols; and the radial distribution. We end by summarizing our findings.

\section{\label{sec:level2}Methods}
\noindent

{\it Initial configurations of Model membrane}\,:\,

We use atomistic molecular dynamics simulations (MD) with {\it GROMACS-2025.1}  \cite{Lindahl} to investigate 1-alkanols $(C_{n}H_{2n+1}OH)$ of varying chain-lengths: $n=2$ (ethanol), $5$ (pentanol), $8$ (octanol), $10$ (decanol), $12$ (dodecanol), $14$ (tetradecanol) and $16$ (hexadecanol) in DOPC  bilayer membrane. {\it PACKMOL} \cite{packmol} is used to build the initial configuration of symmetric DOPC lipid bilayer membrane of $256$ lipids in each leaflet (with a total $512$ lipids) and $16384$ water molecules (such that the ratio of water to lipid is $32:1$) to hydrate the bilayer membrane. We add $128$ number ($25\%$ of total number of lipids) of 1-alkanols with $n=2$, $5$, $8$, $10$, $12$, $14$, $16$ uniformly in both sides of the water layers of the symmetric bilayer membrane, respectively and study total 8 bilayer membranes without and with 1-alkanols, respectively.

\noindent

{\it Force fields}\,:\,

The force field parameters of DOPC are taken from the previously validated united-atom description  \cite{kindt_dopc_dppc_chol,anirban_jpcb12,anirban_cpl13,Tieleman-POPC,mikko}. We have taken the same previously used force-field parameters for the 1-alkanol of different chain-lengths   \cite{bockmann_alkanol,anirban_cpl13,MikkoBPJ2006,ramon_jpcb2011,ramon_plosone2013}. In our simulation study, we use the improved extended simple point charge (SPC/E) model to simulate water molecules, with an extra average polarization correction to the potential energy function.    

\noindent

{\it Choice of ensembles and equilibration}\,:\,

We simulate all 8 symmetric bilayers in the NVT ensemble for $50$\,ps using a Langevin thermostat to avoid bad contacts caused by steric constraints, and then in the NPT ensemble for $1000$\,ns ($1\mu s$) ($T = 296$\,K ($23^{\circ}$ C), $P = 1$\,atm). To collect enough data, we repeat the simulations $4$ times (for a total of $4 \mu s$) for each system. We run the simulations in the NPT ensemble for the first $100$\,ns using Berendsen thermostat and Berendsen barostat, then remaining simulations using Nose-Hoover thermostat and the Parrinello-Rahman barostat to produce the correct ensemble using a semi-isotropic pressure coupling with the compressibility of $4.5\times 10^{-5}$ bar$^{-1}$ for the simulations in the NPT ensemble, long-range electrostatic interactions  by the reaction-field method with cut-off $r_c = 2$\,nm, and the Lennard-Jones interactions using a cut-off of $1$\,nm
\cite{anirban_jcb14,anirban_jpcb12,mikko,patra2004,anirban_cell15}.

\section{Results and discussion}

To confirm that we have attained a thermally and chemically equilibrated bilayer membrane, we calculate the time evolution of the total energy of the system and the area of the lipid (shown in Fig. S1 in Supplementary Information (SI)), exhibiting asymptotic behavior with a very small fluctuation assuring an equilibrated bilayer membrane \cite{anirban_jcb14,anirban_jpcb12,anirban_cpl13}.

We begin our simulation with the initial configuration, in which lipids are homogeneously mixed and 1-alkanols are in the aqueous layers of both sides of the bilayer membrane, and simulate for $1\, \mu s$ to obtain an equilibrated patch of the bilayer membrane, which we analyse to determine 1-alkanol and membrane properties. 

\subsection{Density of 1-alkanol in membrane}

\begin{figure*}[h!t]
\begin{center}
\includegraphics[width=16.0cm]{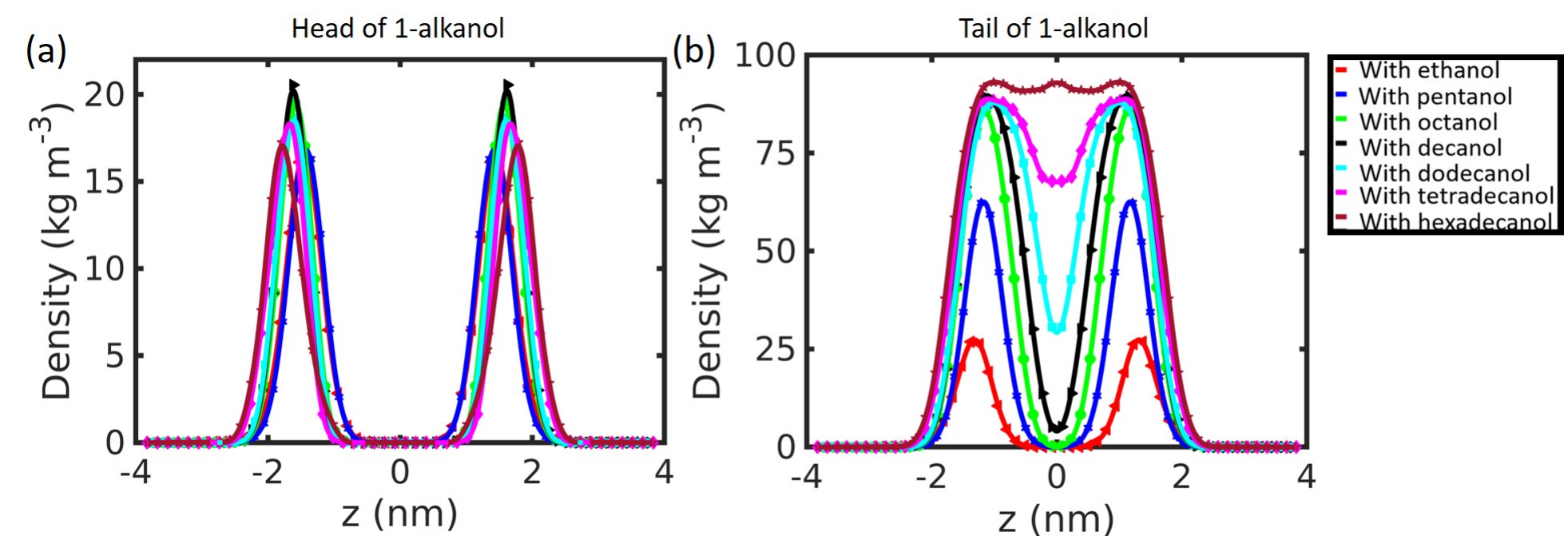}
\caption{The density profile of the head and tail of 1-alkanol in the DOPC bilayer, respectively are shown, where head group comprises with $-OH$ and tail group comprises with acyl chain,  $-(CH2)_{n-1} -CH_3$ with varying $n=2$, $5$, $8$, $10$, $12$, $14$, $16$, respectively.
}
\label{fig1}
\end{center}
\end{figure*}

The density profiles of the head group and the tail of the 1-alkanol for varying chain lengths have been shown in Fig \ref{fig1} (a) and (b), respectively. 
 The atoms of the 1-alkanol's $-OH$ group are considered the head group, while the atoms in the acyl chains, $-(CH2)_{n-1} -CH_3$ with $n = 2$,  $5$, $8$, $10$, $12$, $14$, $16$ are considered the 1-alkanol's tail. As in our simulation, $z-asix$ represents the normal to the surface of the membrane lying in the $xy$-plane, so the densities of the head and tail of the 1-alkanol are plotted on the z-axis. The density profiles of DOPC lipids without and with 1-alkanols are shown in Fig. S2 in SI. The density profile of the head group of 1-alkanol in Fig. \ref{fig1} (a) shows that two peaks of the leaflets are exhibited at the same z-values in both leaflets, respectively. The density profile of the tail of 1-alkanol in Fig. \ref{fig1} (b) shows that two peaks are getting closer with the increasing acyl chain length.  
{\it As a result, we find that the penetration of 1-alkanol increases with the length of acyl chain of the 1-alkanol, $n$. }

\subsection{Deuterium order of DOPC acyl chain}

\begin{figure*}[h!t]
\begin{center}
\includegraphics[width=16.0cm]{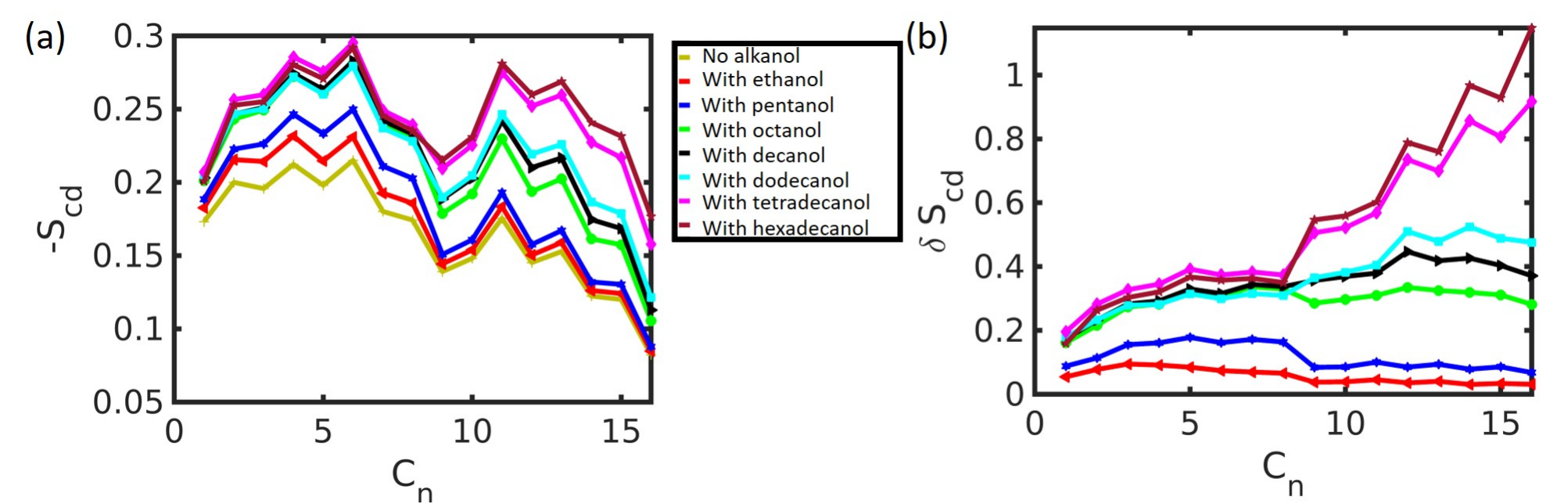}
\caption{The deuterium order parameter, $S_{cd}$ and the fraction of the deuterium order parameter, $\delta S_{cd}$ with the carbon number, $C_n$ of the acyl chain of the DOPC of the bilayer membrane with 1-alkanol with varying acyl chain length, respectively are shown.
}
\label{fig2}
\end{center}
\end{figure*}

The deuterium order parameter, $S_{cd}$, of the acyl chains of the DOPC lipid in the membrane without and with 1-alkanol is calculated as $S_{cd}=\langle \frac{3}{2}(cos^{2}\theta)-\frac{1}{2} \rangle$ where $\theta$ is the angle between the carbon atom and the hydrogen (deuterium) atom and the bilayer normal, as shown in Fig. \ref{fig2} (a).  To show how 1-alkanols affect $S_{cd}$ of the DOPC tails, we define $\delta S_{cd}=\frac{S_{cd}^{0}-S_{cd}^{A}}{S_{cd}^{0}}$ where $S_{cd}^{A}$ and $S_{cd}^{0}$ are the deuterium order parameter of the lipid chains, $C_n$ of the DOPC-lipid bilayer membrane with and  without 1-alkanol, respectively shown in Fig. \ref{fig2} (b).  
There is a prominent deep in the value of $S_{cd}$ at $C_9$ for all membranes without or with 1-alkanol shown in Fig. \ref{fig2} (a) due to the double bond in the DOPC-lipid at $9^{th}$ position of the acyl chain.  Fig. \ref{fig2} (b) shows that the value of $\delta S_{cd}$ with $C_n$ increases with longer acyl chain of the 1-alkanol.
{\it Therefore, we find that the deuterium order parameter ($S_{cd}$) of the DOPC acyl chain decreases with chain length, $n$ of the 1-alkanol acyl chain. }

\subsection{Penetration of 1-alkanol}

\begin{figure*}[h!t]
\begin{center}
\includegraphics[width=16.0cm]{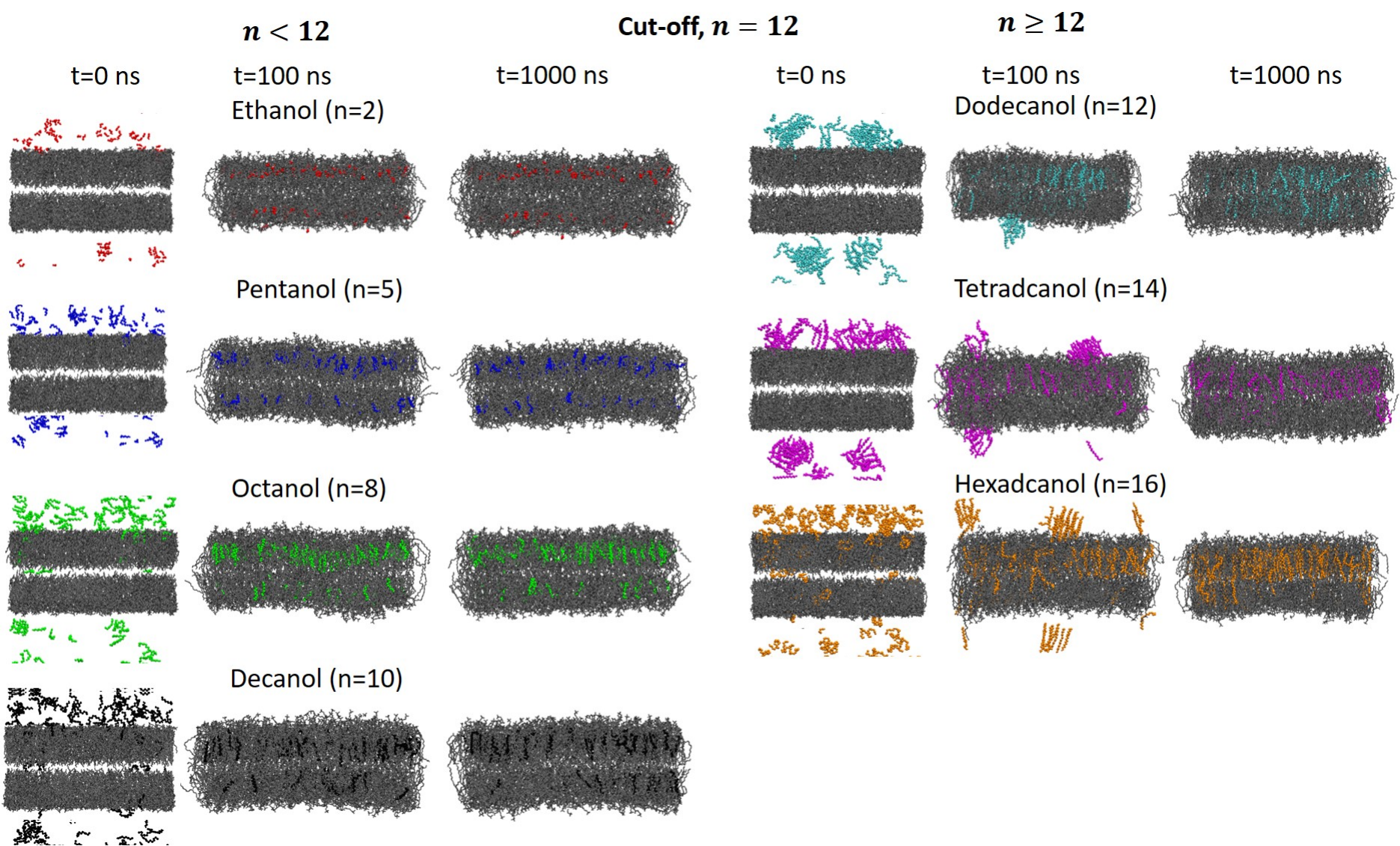}
\caption{The snapshot of the DOPC (grey) bilayer membrane with 1-alkanol with varying acyl chain length, $n=2$ (red), $5$ (blue), $8$ (green), $10$ (black), $12$ (cyan), $14$ (magenta), and $16$ (orange) at different time, $t=0 \, ns$,  $t=100 \, ns$, and  $t=1000 \, ns$, respectively are shown.
}
\label{fig3}
\end{center}
\end{figure*}

\begin{figure*}[h!t]
\begin{center}
\includegraphics[width=19.0cm]{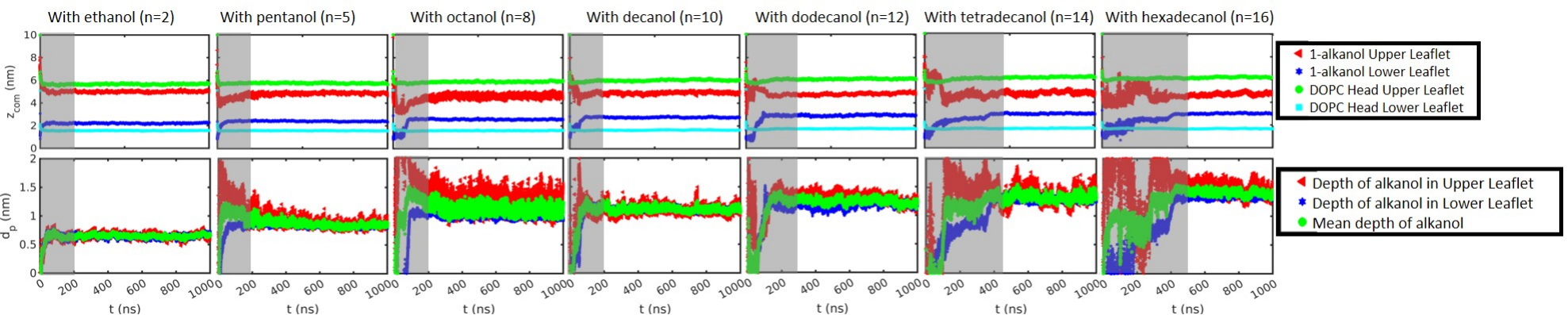}
\caption{The time evaluation of the $z$-component of the center of mass, $z_{com}$ and the depth of penetration, $d_p$ of the 1-alkanol in the DOPC bilayetr membrane with  1-alkanol with varying acyl chain, respectively are shown.
}
\label{fig4}
\end{center}
\end{figure*}

\begin{figure*}[h!t]
\begin{center}
\includegraphics[width=16.0cm]{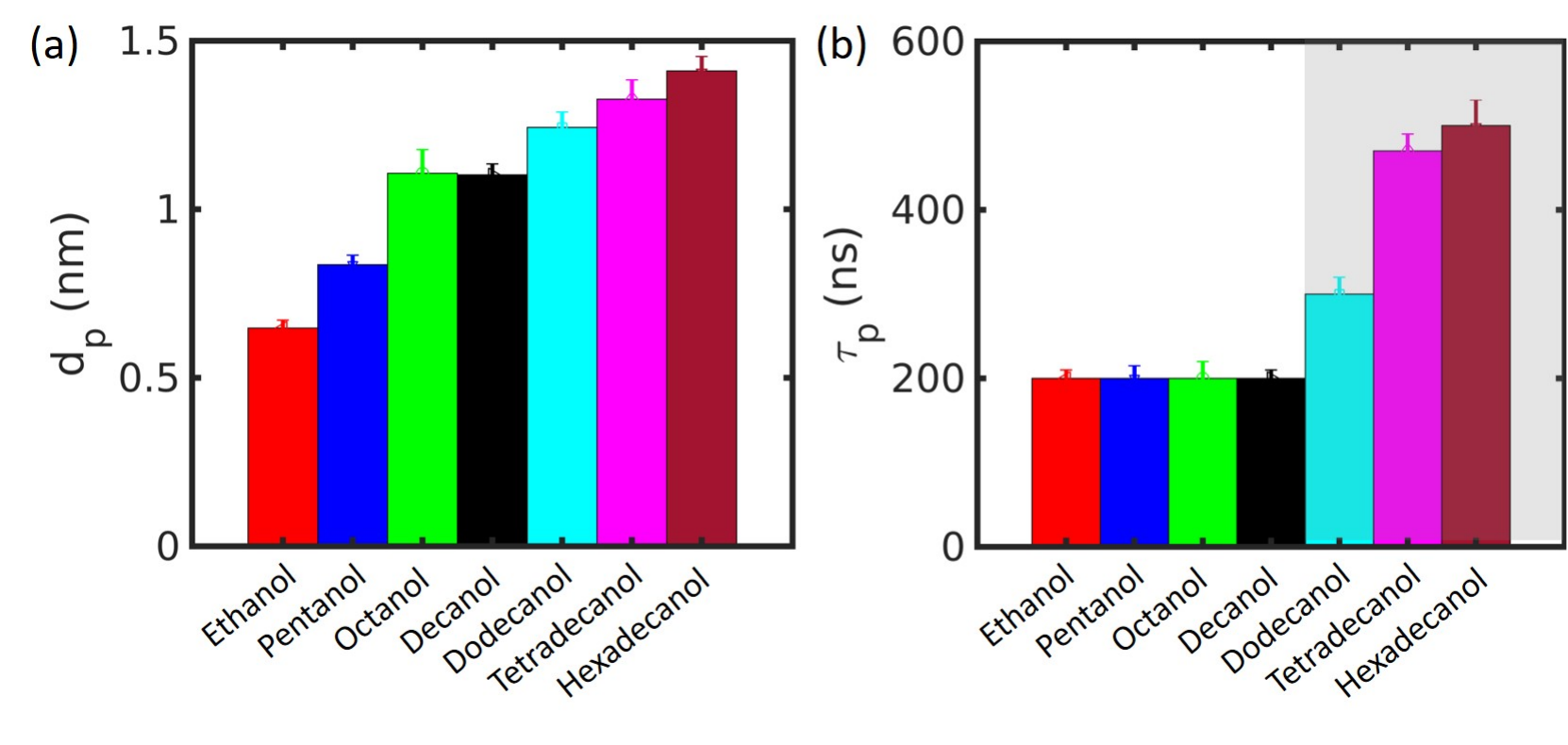}
\caption{The depth of penetration, $d_p$ and the time of penetration, $\tau_p$ of the 1-alkanol in the DOPC bilayer membrane with 1-alkanol with varying acyl chain length, respectively are shown.
}
\label{fig5}
\end{center}
\end{figure*}

Fig. \ref{fig3} shows the snapshot of 1-alkanol with varying chain length in the membrane at time,  $t=0 \,ns$, $t=100 \,ns$, and $t=1000 \,ns$, respectively which shows how the 1-alkanol with varying chain length differs in the penetration of the membrane. We collect the coordinates of the atoms in the head group of DOPC, entire DOPC lipids, and 1-alkanol molecules with time from the simulation of $1\, \mu s$ of DOPC with/ without 1-alkanols. 
%
%
We begin by calculating the lipid membrane's centre of mass (COM) ($x_{com}^{membrane}$, $y_{com}^{membrane}$, $z_{com}^{membrane}$) for each time frame.
We also calculate the COM ($x_{com}$, $y_{com}$, $z_{com}$) of the head group of each lipid and that of the entire 1-alkanol molecule with time, respectively. 
 With time, we form groups of DOPC lipids and 1-alkanol molecules that remain in the upper ($z_{com}>z_{com}^{membrane}$) and lower ($z_{com}<z_{com}^{membrane}$) leaflets, respectively. 
Now, we calculate the average COM of the head group of DOPC lipid and 1-alkanol of the upper and lower leaflet, respectively, averaged over the number of lipids/ 1-alkanol in each leaflet. The time evaluation of the z-component of the COM ($z_{com}$) of the head group of DOPC lipid and 1-alkanol of the upper and lower leaflet is shown in Fig. \ref{fig4}.
The depth of the penetration, $d_p$ of the 1-alkanol is defined as, $d_p^{upper(lower)}=|(z_{com}^{upper(lower)-dopc-head-group}-z_{com}^{upper(lower)-alkanol})|$.
As shown in Fig. \ref{fig5} (a), the penetration depth of 1-alkanols, $d_p$, increases with the value of acyl chain length, $n$. The time evaluation of the depth of penetration of 1-alkanol in the upper and lower leaflets and the mean value of them are shown in Fig. \ref{fig4}, where we highlight in color grey the value $d_p$ of 1-alkanol in the bilayer membrane until it reaches the asymptotic values, which is denoted as the penetration time, $\tau_p$ of 1-alkanol.
Fig. \ref{fig5} (b) shows that 
the time of penetration of 1-alkanols with different acyl chains has a cut-off value, $n=12$, below which 1-alkanol molecules can penetrate to the membrane within $100-200\, ns$, but 1-alkanols with longer acyl chains take a long time (approximately $300-500\, ns$). 
{\it As a result, we find that the penetration time of 1-alkanol has a cut-off value ($n = 12$) for the acyl chain of 1-alkanol, even though the depth of penetration of 1-alkanol increases with acyl chain length.}

\begin{figure*}[h!t]
\begin{center}
\includegraphics[width=16.0cm]{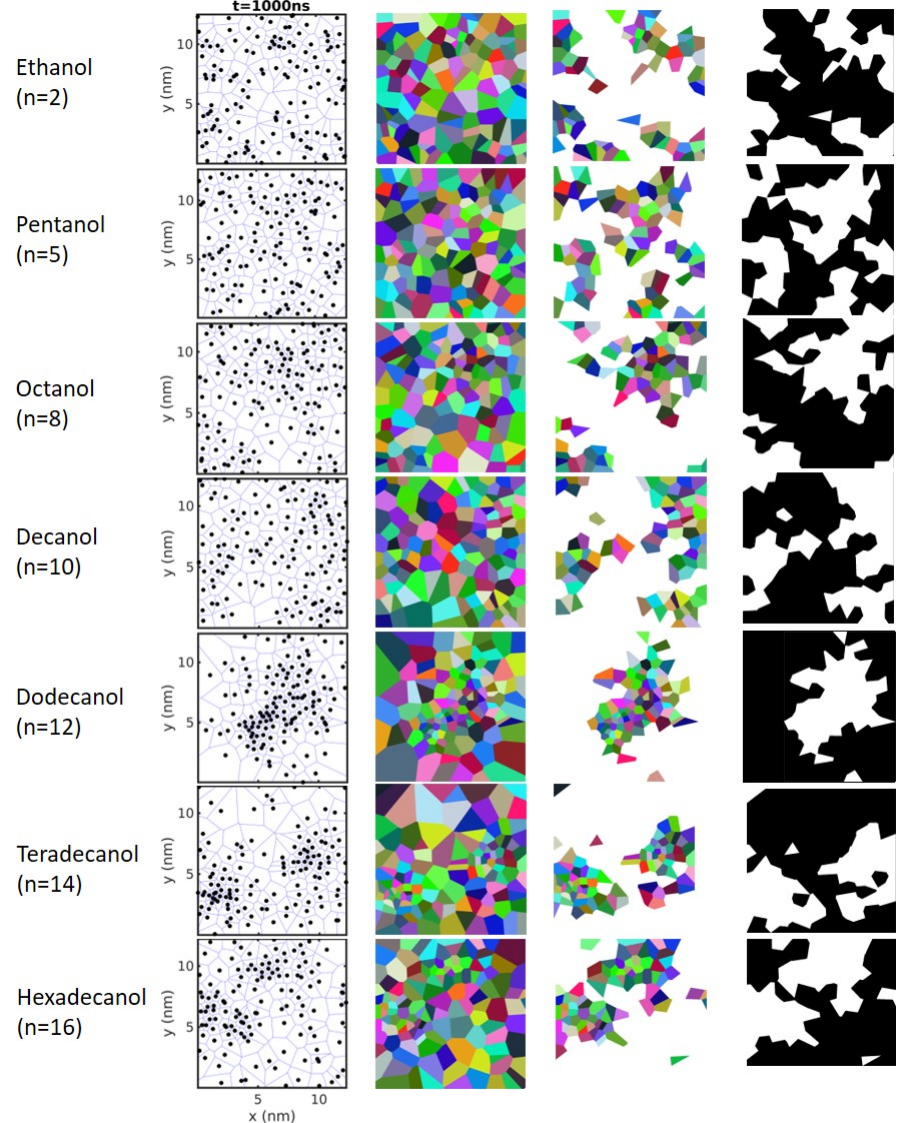}
	\caption{The clusterization of 1-alkanols using Voronoi Tessellation technique are shown for the 1-alkanol (black circle) in the DOPC bilayer membrane with varying acyl chain length, respectively are shown.
}
\label{fig6}
\end{center}
\end{figure*}

\begin{figure*}[h!t]
\begin{center}
\includegraphics[width=18.0cm]{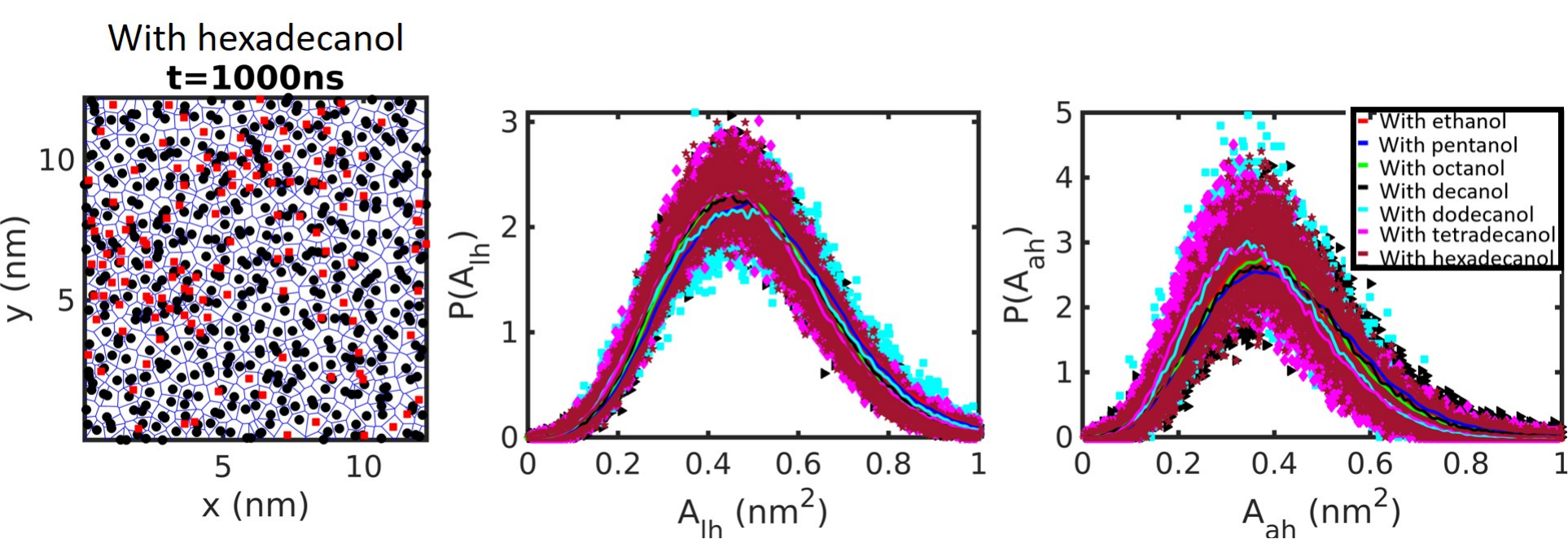}
	\caption{The Voronoi diagram of the center of mass of the hexanol (red square) and  DOPC lipid (black circle) together at time, $t=1000\, ns$ and the probability distribution of the area per lipid and 1-alkanol for the bilayer membrane with  1-alkanol with varying acyl chain, respectively are shown.
}
\label{fig7}
\end{center}
\end{figure*}

\begin{figure*}[h!t]
\begin{center}
\includegraphics[width=18.0cm]{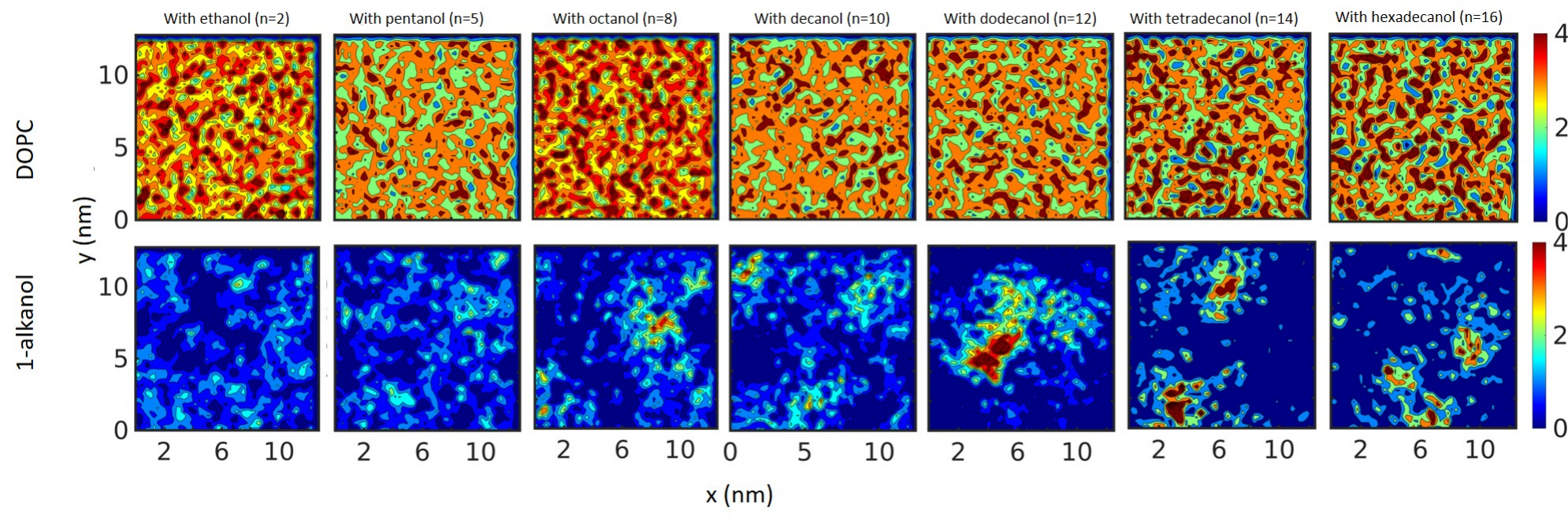}
\caption{The number density of the DOPC and 1-alkanol averaged over last $500\, ns$ trajectory of the entire simulation of $1\, \mu s$ of  the DOPC bilayetr membrane with  1-alkanol with varying acyl chain, respectively are shown.
}
\label{fig8}
\end{center}
\end{figure*}

\subsection{Clusterization of 1-alkanols}
To study the clusterization of 1-alkanol in the lipid membrane, we have developed a novel technique based on Matlab-image analysis tool fusion with the Voronoi tessellation technique. 
Our method can be widely applicable to the study of clusters of molecules in any molecular dynamics \cite{anirban_jpcb12,anirban_cell15,anirban_jcb14} or Monte Carlo simulation \cite{anirban_prl16} as well as any experiments describing spatial heterogeneity of molecules such as fluorescent-based experiments (confocal FM, FRET, TIRF) \cite{anirban_cell15} to study the spatial organisation of cell membrane components.


We use last $500\, ns$ trajectory out of the entire $1000\, ns$ simulation of each system, where we have saved each frame in a $100\, ps$ interval during our simulation, and a total of $5000$ frames of trajectory have been used to study clusterization of 1-alkanol. We collect the (x,y) positional coordinates of all 1-alkanol molecules for each frame and calculate the COM of each 1-alkanol molecule. The steps 
to calculate the cluster of 1-alkanol in each frame are given below.


First, we plot the $x$ and $y$-components of the COM of each 1-alkanol 
for each frame (For example, here, the spatial distribution of COM of each 1-alkanol is shown at time-frame $t = 1000\, ns$ in the left-most column figure in Fig. \ref{fig6}.
Second, we use Voronoi tessellation 
technique to obtain Voronoi-diagram of 1-alkanol
 and highlight the Voronoi-cells with different colors corresponding to each 1-alkanol, as shown in the second column figure in Fig. \ref{fig6}.
 Third, from the Voronoi-cells of the 1-alkanol, we calculate the density by inverting the area of each cell, as each Voronoi-cell contains one 1-alkanol molecule, and we calculate the mean density of the 1-alkanol. Now, only Voronoi-cells with densities greater than the mean density are selected and
  the Voronoi-cells in different colors with high densities 
  are shown in the third column figure in Fig. \ref{fig6}.
   Fourth, as shown in the fourth column of Fig. \ref{fig6},
   we fill-up all Voronoi-cells with higher density
   in white color while the regions of Voronoi-cells with lower density are shaded in black color, respectively and save them in high resolution ($\sim 1024p$). We notice that the white region in the figure is a cluster of 1-alkanol. Now, we read the high-resolution black-and-white image of the cluster in Matlab and use the function `bwlabel' to calculate the area of the polygonal white space representing the cluster based on the pixels as a smallest bin-size of the image.


From all the frames in the last $500\, ns$ trajectory of the simulation, we gather all the values of the area of clusters of 1-alkanol exhibited in each frame. We plot the probability distribution of the area of the cluster, $P(A_{cluster})$, shown in Fig. \ref{fig7},
which shows the formation of a distinct cluster only for 1-alkanol with an acyl chain ($n = 12$, $14$, and $16$) at $20\, nm^2$, $30\, nm^2$, and $45\, nm^2$, respectively. Thus, 1-alkanols with higher acyl-chain lengths ($n \geq 12$) exhibit prominent clusterization in lipid membranes, and there is no cluster exhibited for 1-alkanols with shorter acyl chains ($n < 12$). However, we need to use Voronoi tessellation on both the lipid and the 1-alkanol in the same frame and calculate the probability distribution of the area of the lipid and the 1-alkanol, and estimate the area as $0.5 \pm 0.1 nm^2$ and $0.38 \pm 0.1 nm^2$ for DOPC and the 1-alkanol, respectively, as shown in Fig. \ref{fig7}.


{\it Therefore, we find that there is a cut-off value, $n = 12$, of the acyl chain length of 1-alkanol above which the hydrophobic effect of the acyl chains is strong enough to form a cluster of 1-alkanol. }

 
We can calculate clusters of 1-alkanol from the spatial number density plot shown in Fig. \ref{fig8} by dividing the system's box with 
$12\times12$ square grids with a
bin-size $1\, nm$ (considering, the size of the membrane is approximately $12 \times 12 \, nm^2$)and then getting the high-density bins, which are denser than average, followed by cluster analysis. Our method, demonstrated above, is efficient compared to 
the square grid method. Using the Voronoi tessellation method, we get the actual polygonal area where the COM of molecules is present at the center and we obtain polygonal clusters of the 1-alkanols. 
Since there is a possibility that the borders of the bins will contain the COM of molecules, if we use the square grid approach to calculate number density, we might run into artefacts. 
 Furthermore, by analyzing a black-and-white image that is saved to define the the cluster of 1-alkanols, we improve our calculation for determining the properties of the cluster, because the smallest bin size in our method is the size of a pixel in the image.

\begin{figure*}[h!t]
\begin{center}
\includegraphics[width=16.0cm]{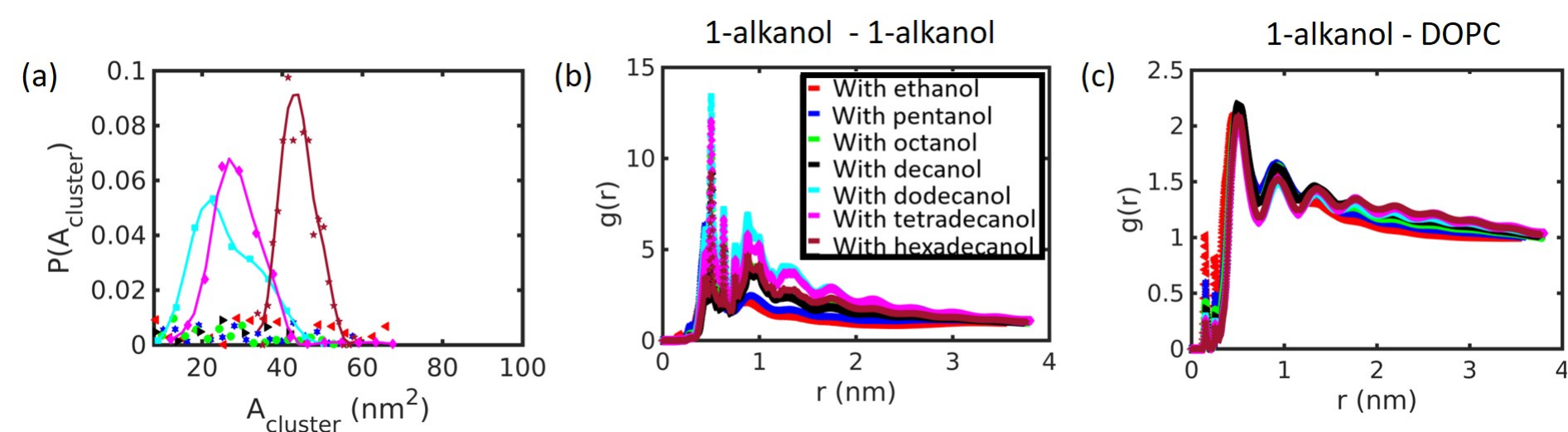}
\caption{The Probability distribution of the area of clusterization, $P(A_{cluster})$ of 1-alkanols are shown in Fig\ref{fig9} (a) for the 1-alkanol in the DOPC bilayer membrane with varying acyl chain length, respectively. The radial distribution, $g(r)$ of the 1-alkanol - 1-alkanol pair and 1-alkanol - DOPC pair for the 1-alkanol in the DOPC bilayer membrane with varying acyl chain length are shown Fig\ref{fig9} (b) and (c), respectively.
}
\label{fig9}
\end{center}
\end{figure*}

\subsection{Radial distribution}

We calculate the radial distribution, $g(r)$, of the pair distance between two 1-alkanols and that of the pair distance between 1-alkanol - DOPC lipid, as shown in Fig.  \ref{fig9} (b) and (c), respectively, for varying acyl chain length, $n$, of 1-alkanol. In dodecanol, tetradecanol, and hexadecanol, there are prominent peaks at $0.5 \, nm$ in the $g(r)$ of the 1-alkanol—1-alkanol pair shown in Fig. \ref{fig9} (b), which are very small in ethanol, pentanol, octanol, and decanol. Because of the strong hydrophobic effect of the long acyl chains, the 1-alkanol with a long acyl chain ($n \geq 12$) interacts with itself and forms clusters. Short acyl chain 1-alkanols, on the other hand, do not form clusters due to low hydrophobic interaction. Fig. \ref{fig9} (c) shows prominent peaks in the $g(r)$ of the 1-alkanol-DOPC pair at $0.25\, nm$ for ethanol, pentanol, octanol, and decanol, respectively indicating comparatively higher interaction with lipids, which are absent for dodecanol, tetradecanol, and hexadecanol. Thus, 1-alkanol with a short chain acyl chain is homogeneously distributed in the membrane, as shown in Fig. \ref{fig6}, and can penetrate the lipid membrane easily due to its small size while it takes a long time for a lump of clustered 1-alkanol to penetrate the membrane.


{\it Therefore, 1-alkanols with a higher acyl chain length ($n \geq 12$) interact together more than that of a lipid, leading to the formation of cluster and take a long penetration time, while 1-alkanols with a lower acyl chain length ($n < 12$) interact with lipids less than that of themselves, are distributed uniformly in the membrane, and take a shorter penetration time.}


\section{Conclusion}

We present the effect of 1-alkanols with different chain lengths in the symmetric bilayer membrane comprising DOPC using atomistic molecular dynamics simulation. Here, we explore the effects of 1-alkanols
with varying chain length in
  the membrane. Our main findings in the present study are the following: the depth of penetration of 1-alkanol into the membrane increases with the chain length (n) of the acyl chain of the 1-alkanol. The deuterium order parameter ($S_{cd}$) of the DOPC acyl chain, $C_n$, increases with the length of the 1-alkanol acyl chain. We find a cut-off value ($n = 12$) of the acyl chain length of the 1-alkanol, above which the membrane exhibits clusters of the 1-alkanol due to higher interaction among themselves. As a result, 1-alkanols with an acyl chain with a higher cut-off value take longer to penetrate the membrane, as the lump of 1-alkanols requires more time to get enough space to penetrate the membrane. 1-alkanols with short chain lengths, on the other hand, are distributed uniformly in the membrane and interact with lipids in the membrane, penetrating the membrane in a short period of time due to their small size. 
  This could explain why 1-alkanol with a shorter acyl-chain is more effective as an anesthetic.
Our findings show that both the penetration depth and time of 1-alkanol molecules into the membrane vary with their acyl chain length
and could be useful for the study of 1-alkanols and their anesthetic effects.


\begin{acknowledgments}
A.P.  gratefully acknowledges the hospitality of generous the computing facilities of  clusters at  SASTRA University, Thanjavur, Tamilnadu. 
A.P. acknowledges the support under Science and Engineering Research Board (SERB), Department of science and technology, Government of India [SERB-SRG/2022/001489] and T.R. Rajagopalan research fund, SASTRA University, India.

\end{acknowledgments}

\bibliography{referencesbilayer_alcohol}
\end{document}